\documentclass[letterpaper, 12pt]{article}

\usepackage{cmjStyle} 
\usepackage{natbib} 
\bibpunct{(}{)}{;}{a}{}{, } 
\raggedright 
\usepackage{epstopdf}
\usepackage{amsmath,amssymb,amsbsy,bm,upgreek,nicefrac}
\usepackage{todonotes,microtype}

\usepackage{spverbatim}
\usepackage{listings}
\usepackage{tcolorbox}
\tcbuselibrary{breakable,skins,listings}

\makeatletter
\tcbset{
    addtolol/.style={list entry={\kvtcb@title},add to list={lol}{lstlisting}},
}
\makeatother

\newtcblisting[use counter=lstlisting]{cmhlisting}[3][]{%
    width=\textwidth,
    breakable,
    colback=green!2!white,
    colframe=white!85!black,
    listing only,
    listing options={#1},
    center title,
    title={\color{black}{\scshape Listing \thetcbcounter}: ~#2},label={#3},
    addtolol,
}

\graphicspath{{./Figures/}}

\begin{document}

{\cmjTitle Large Language Models: From Notes to Musical Form}
\vspace*{24pt}

{\cmjAuthor Lilac Atassi}	
 
\begin{cmjAuthorAddress}
	Department of Music\\
        University of California San Diego\\
        La Jolla, CA 92093-0099\\		
	latassi@ucsd.com
\end{cmjAuthorAddress}

\begin{abstract}
While many topics of the learning-based approach to automated music generation are under active research, musical form is under-researched. In particular, recent methods based on deep learning models generate music that, at the largest time scale, lacks any structure. In practice, music longer than one minute generated by such models is either unpleasantly repetitive or directionless.

Adapting a recent music generation model, this paper proposes a novel method to generate music with form. The experimental results show that the proposed method can generate 2.5-minute-long music that is considered as pleasant as the music used to train the model. The paper first reviews a recent music generation method based on language models (transformer architecture). We discuss why learning musical form by such models is infeasible. Then we discuss our proposed method and the experiments.
\end{abstract}

\section{Introduction}
With the most recent generation of generative machine learning (ML) models, a revived interest in ML-based music generation methods has appeared. Generative Adversarial Networks (GANs) \citep{goodfellow2014generative} and Variational Autoencoders (VAEs) \citep{kingma2014auto} were among the earliest in this recent wave of new generative models. The computation cost of GANs and VAEs is prohibitive for learning long sequences like text. This is mostly due to the computational cost of convolution filters. The transformer architecture \citep{vaswani2017attention} led to multiple approaches to generative models for text. The attention layer in the transformer architecture proved to be more efficient than convolution filters for learning the relationships in long sequences. 

The original transformer model is adapted by two prominent models. The first approach is an encoder-only architecture, which is used by Bert \citep{devlin2018bert} to encode text for classification tasks. Such models are trained to estimate $P(t_k|t_0,\dots,t_{k-1},t_{k+1},\dots,t_n)$, probability of the input at index k given the rest of the input. The second approach is a decoder-only architecture which is used by GPT \citep{achiam2023gpt} to generate text. The model is trained to estimate $P(t_k|t_0,\dots,t_{k-1})$, the probability distribution of the input at the last index given the prior indices. This autoregressive model is simpler than the model used by Bert, requiring fewer parameters. Consequently, the less computationally complex model of GPT can be designed to process longer input sequences. In the GPT family of models, GPT-1 to GPT-3, the input sequence length is increased and the models are deeper, which, combined with larger datasets, leads to improved performance. An initial test by \cite{brown2020language} on people's ability to tell whether a 500-word article was written by humans or GPT-3 show a mean accuracy of 52\%, just slightly better than random guessing.

All the new wave of generative models have been explored in the literature to generate music. Jukebox \citep{dhariwal2020jukebox} is based on Hierarchical VQ-VAEs \citep{razavi2019generating} to generate multiple minutes of music. This is one of earliest models that could generate longer than one minute of music. Notably, the authors mention that the generated music at a small scale of multiple seconds is coherent, and at a larger scale, it lacks musical form.

Music Transformer \citep{huang2018music} adapted the transformer architecture for music generation, capable of processing the number of input tokens equivalent to up to one minute of music. The generated music sometimes would show structures that are musically coherent up to the scale of about one minute. The model can be trained on symbolic music or audio. More recent works on transformers and autoregressive models first use an efficient encoder to compress the audio. In the compressed space, the transformer model is trained to generate new tokens. Then, the generated tokens are converted into audio by a decoder. MusicGen \citep{copet2024simple} follows this approach. It uses EnCodec \citep{defossez2022high} to encode and decode the audio. After the MusicGen model is trained, it generates token in the latent space of EnCodec, and the decoder is used to convert the tokens into audio.

The generative models in the literature have been applied to both symbolic and audio music. Most of the recent larger models are trained on audio, as collecting a large training set of music audio is more readily than symbolic music. Another argument is that generating audio directly can be more expressive than using a synthesizer to convert generated symbolic music into audio.

Earlier works on generative music models, such as JukeBox, attempted to optimize the model to process long sequences of multiple minutes in the hope that musical structures and form at all scales are learned by the model. However, none of the models in the literature have demonstrated musical structure at a large scale and lack even simple musical forms. We argue this is a principled limitation due to the nature of the data, as discussed in the following section. Using Large Language Models (LLMs) we propose an approach to introduce musical form in generated music.

\parskip 18pt

%
\section{Unlearnable Musical Form}

To illustrate the problem that generative models struggle with learning musical form, consider a simple case. A generative model using maximum likelihood estimation optimize $\theta$ to estimate the joint probability $p_{\theta}(t_1,t_2)$ from the data samples with discrete and finite values. A parametric model can estimate the joint probability if the training data samples are on a compact manifold. If there is a large amount of variation in $t_1$ in the samples, then $p_{\theta}(t_1,t_2)$ is reduced to $p_{\theta}(t_2)$ as $p(t_1)$ becomes virtually uniform. With high dimensional data, the problem of non-compact data manifold is compounded with the fact that with an arbitrarily good model as judged by average log-likelihood, $-log[p_{\theta}(t_1,\dots,t_d)] \approx d$. The optimization process minimizes the negative log-likelihood by adjusting $\theta$ for the dimensions with less variation. In a general musical corpus, with many parameters varying across music pieces with the same musical form, the amount of variation at large temporal scales to learn musical form is large enough that even simple musical forms are impossibly difficult to learn for generative models.

\begin{figure}[!ht]
\begin{center}
\includegraphics[width=0.55\textwidth]{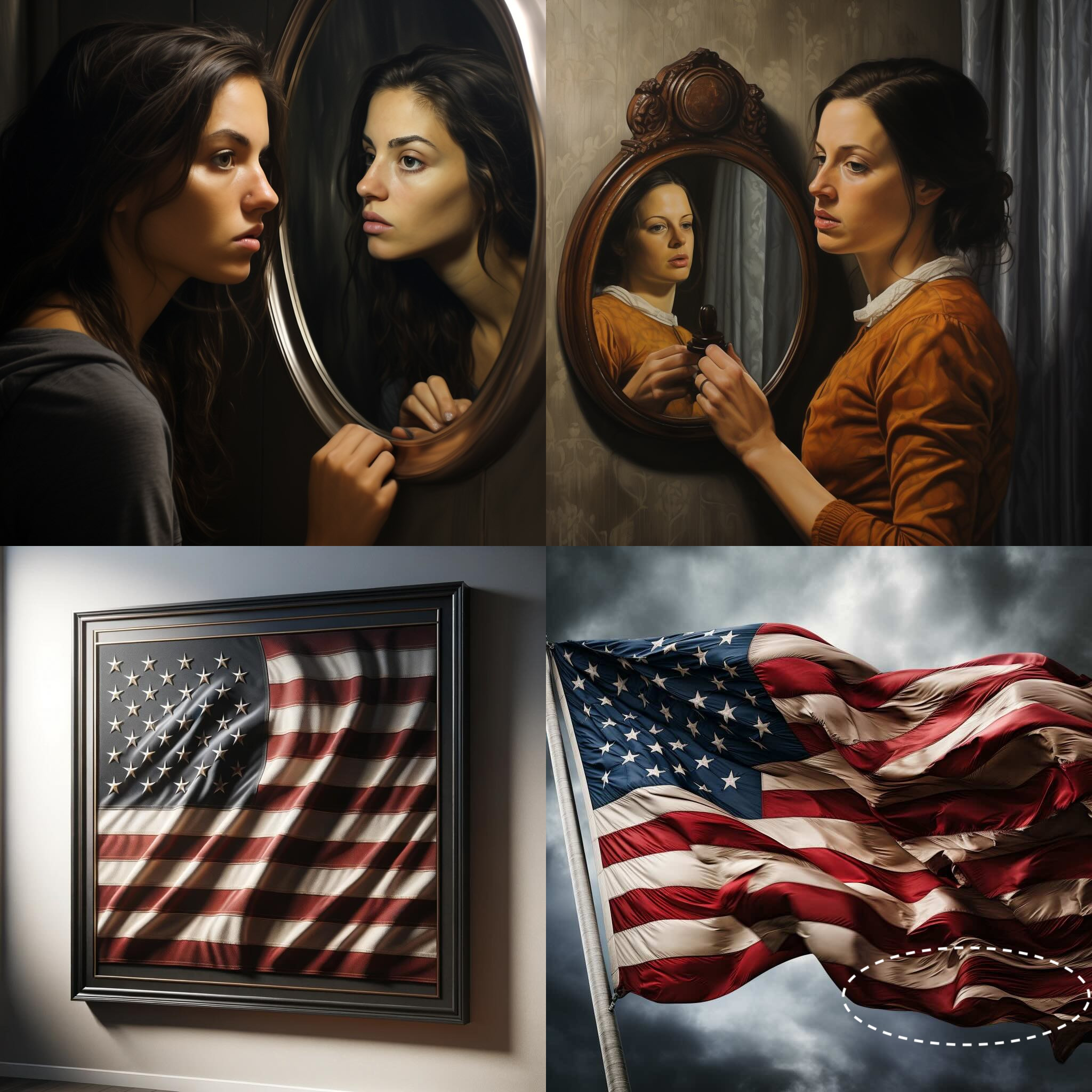}
\caption{Illustrating the incoherence in images generated by Dall-E, the top row images demonstrate that the generative model has learned that a mirror reflects an image, yet the incoherence in the generated images is evident. The bottom left image shows that the model can generate flags with coherence at all scales. However, the bottom right image reveals the model's struggle with coherence at large scales when generating wavering flags, with one of the discontinuities highlighted.}
\label{fig:vis_form}
\end{center}
\end{figure}

The problem of large variation is not unique to music data. Image generative models, such as those for drawing hands, struggle due to the extensive variability; in most training images, some fingers are occluded, leading the model to fail in learning the correct number of fingers a hand should have. Additionally, the object occluding the fingers can vary significantly, from fingers occluding each other to coffee cups, pockets, torsos, and other objects, preventing the model from learning to draw an occluding object for hands with fewer than five fingers. This issue extends to other structures with a high degree of variation in images. Figure \ref{fig:vis_form} illustrates two other structures with significant variation that Dall-E fails to learn: the angle of the subject, mirrors, and points of view vary enough that the model cannot learn how a coherent reflected image should appear. Wavering flags present a similar challenge; due to the variation at large spatial scales in the training images, the model generates physically implausible images.

Our work builds on MusicGen, a generative music model that relies on EnCodec to compress audio. The Following section briefly reviews these two methods.

\section{EnCodec and MusicGen}

The most recent methods in the literature for music generation follow the approach of Stable Diffusion \citep{rombach2022high}. In that, the generative model is trained in the latent space of an encoder. The generated vector is then decoded into audio using the corresponding decoder. The encoder and decoder are trained on a separate dataset prior to training the generative model. In the audio domain the most commonly used in the literature include EnCodec and soundStream \citep{zeghidour2021soundstream}. The main reason for training the generative model in the compressed latent is to reduce the computational cost of generating long audio. EnCodec, for instance, offers a compression ratio of 150:1. Some of the earlier work following this approach were based on WaveNet \citep{van2016wavenet} and AutoEncoders \citep{kingma2014auto}. 

The encoder of EnCodec is a convolution network. The architecture is similar to other common models, with residual connections, downsampling through strided convolution, and doubling the number of convolution channels whenever downsampling occurs. The convolutions blocks are followed by a two-layer long short-term memory (LSTM) to provide sequence modeling. The output of the LSTM is then passed to the last 1D convolution layer with a kernel size of 7 and 128 output channels to generate the latent vector. The architecture and parameters are set such that the encoder outputs 75 latent steps per second of audio at 24 kHz, and 150 at 48 kHz. There are variants of the model with a small difference in some parameters, a non-streamable or high-fidelity model, and a streamable model.  For the non-streamble model, the input is split into chunks of 1 seconds, with an overlap of 10 ms. The streamable model, after receiving 320 samples (13 ms), outputs 320 samples (13 ms).

The decoder part of the model, similar to any autoencoder, mirrors the encoder part. To upsample the feature maps in the decoder, transposed convolution is used. The decoder's output is either mono or stereo. The whole system is trained end-to-end to minimize a reconstruction loss applied over both the time and frequency domains, together with a perceptual loss in the form of discriminators operating at different resolutions.

The latent vectors are quantized for further compression. Vector quantization involves projecting an input vector onto the closest entry in a codebook of a given size. Residual vector quantization refines this process by computing the residual after quantization and further quantizing it using a second codebook, and so forth. EnCodec has 1024 entries in each codebook, equivalent to 10 bits per codebook. The codebooks are updated during the training process and are frozen during the inference process. Multiple variants of EnCodec are trained with a varying number of codebooks. For instance, with four codebooks, the first one is used to quantize the latent vectors directly. The second codebook is used to quantize the residual or error from the first quantization, and the third and fourth codebooks are used to quantize the residuals from the previous quantization. To convert the discrete representation back to a vector, the corresponding codebook entries are summed before going into the decoder.

In MusicGen, an autoregressive model is trained in the quantized latent space of EnCodec to model music. Two variants of MusicGen are trained on 10 and 30 seconds of music audio. Given that the latent vectors are quantized, the problem of modeling sequences of them as sequences of tokens can be treated similarly to modeling sequences of characters in text. Therefore, MusicGen, similar to a natural language model, learns from sequences of tokens. The only difficulty is that each one of EnCodec's vectors is quantized to multiple, for instance, four, tokens. And the second token depends on the first token as it is computed from the quantization error of the first token.

In MusicGen, a general approach is proposed to generate the tokens, experimenting with multiple patterns of prediction. Both subjective and objective evaluations show that the exact and slow flattening patterns outperform the others. The second-best performing pattern in terms of generated music quality is the delay pattern, which predicts four tokens at each step. In the delay pattern, at step t, the model predicts the token from the first codebook at t, the token from the second codebook at t-1, the token from the third codebook at t-2, and so on.

To condition the model on text, multiple text encoders are evaluated, with T5 \citep{raffel2020exploring} shown to have the best subjective and objective music quality in the experiments. During training, the text condition is dropped with a probability of 0.2. Classifier-free guidance (CFG) is used to generate samples with text conditioning. First, the probability distributions for the tokens are predicted without text conditioning. In the second step, the probability distributions for the tokens are predicted with text conditioning. Then, the two probability distributions are interpolated at each step, and a sample is taken from each interpolated distribution.

MusicGen is trained on 20K hours of music. Half of the training data is private and internal at Meta. The other 10K hours of training music are taken from ShutterStock (\url{https://www.shutterstock.com/music}) and Pond5 (\url{https://www.pond5.com}) vocals-free music data collections.

\section{Controlling MusicGen by a Language Model}

As MusicGen is conditioned on text, it affords an interface in natural language. Thus, an LLM can generate prompts for MusicGen to replace human prompts. Generally, an LLM needs to be prepared for a specific task, such as generating prompts that MusicGen can understand. There are two approaches to adapting an LLM for a task. Fine-tuning or training a pre-trained LLM is one approach. LoRA \citep{hu2021lora}, for example, is an efficient method to train an LLM on a small dataset for a specific task. Another approach is in-context learning through prompt design. Prompts are typically composed of a task description and some canonical examples \citep{brown2020language}. In-context learning has the advantage of requiring far fewer samples and being much less resource-intensive for trying and evaluating several prompts.

The LLM is instructed to generate prompts for each part of the music piece. Switching between prompts passed to MusicGen creates a sudden jump. To ensure a smooth transition between parts, we modify the CFG method. Instead of estimating one conditioned probability distribution, we estimate two conditioned distributions. In the interpolated probability distribution, the weight of the first conditioned distribution over five seconds goes from one to zero linearly, and the weight of the second conditioned distribution increases from zero to one. This method allows the model to generate tokens that facilitate a continuous transition between parts.

Similarly, to generate parts with variations of the prior parts, we provide a 15-second audio prompt from a previous part to MusicGen. This 15-second audio is passed to the encoder of EnCodec to generate the corresponding tokens. Then, the prompt tokens are passed to the predictive model of MusicGen. The weight of the probability distribution over tokens, conditioned on these audio prompt tokens, is then linearly reduced to zero over 10 seconds.

We use ChatGPT 4 which is GPT 4 \citep{achiam2023gpt} fine tuned to answer questions. We found that providing a diverse set of sample MusicGen prompts to ChatGPT helps in generating diverse prompts by ChatGPT. Therefore, we include several music descriptions from Pond5 in the prompt. Asking ChatGPT to first come up with a musical form and then write prompts for each part aids in achieving diversity in the musical forms. The instruction prompt also specifies that the length of each piece is 2.5 minutes. Our instruction prompt to ChatGPT is presented in Listing \ref{pgpt}.

\begin{cmhlisting}{Prompt to ChatGPT}{pgpt}
Assume you're a musician. You can write text prompts for a system 
that generates the music for the given description of the music. The system was trained with description from a stock music catalog, descriptions that will work best should include some level of details on the instruments present, along with some intended use case (e.g. adding "perfect for a commercial" can somehow help). 
The following are some example prompts that the system understands and the music it can generate: 
- "A cheerful country song with acoustic guitars" 
- "Lofi slow bpm electro chill with organic samples" 
- "Electro swing" 
- "90s rock song with electric guitar and heavy drums" 
- "a light and cheerly EDM track, with syncopated drums, aery pads, and strong emotions bpm: 130" 
- "Chillstep, calm EDM" 
- "A smooth jazz fusion piece blending electric piano, a stand-up bass, and a soft trumpet solo. Perfect for a sophisticated lounge setting" 
- "A relaxing tropical house track with steel drums, a gentle electronic beat, and a breezy melody, great for a summer beach party scene" 
- "Complextro" 
- "Smooth guitar ballad" 
- "An ambient music piece with ethereal synth pads, a gentle piano melody, and subtle nature sounds, creating a serene and mystical atmosphere" 
- "Japanese traditional music koto" 
- "Japanese traditional music flute and koto" 
- "J-Pop, 50bpm" - "A classic soul track with a warm bass line, smooth electric guitar, and soulful vocals, capturing the essence of a 70s soul club." 
- "A lively Latin track with vibrant brass, energetic percussion, and catchy piano riffs, suitable for a festive party or dance scene" 
- "A nostalgic 80s synthwave track with vintage synthesizers, a driving bass line, and electronic drums, perfect for a retro-themed project or game." 
- "A sweeping orchestral composition with soaring strings, powerful brass, and dramatic percussion, ideal for an epic movie trailer" 
Don't limit yourself to these sample prompts. Be creative and compose music piece by writing prompts for each part. Such that after stitching the parts together an impressive music piece is generated that most people would find creative and enjoyable. The music piece should be coherent and have unity. Write the thought process for the composition, followed by the parts. Don't change the music style often so the piece is unified and holds cohesion. Consider complex music form for the piece and describe the form first. Very important constraints that your prompts have to satisfy: 1. The whole piece should be exactly **150 seconds**. You also decide how long each part should be. 2. The prompt of each part can reference another part. REFRENCED_PART is either the number of a previous part or -1 meaning no other part is referenced. 3. Consider David Huron's ITPRA framework when planning the music form. 4. Do not use comparative adjectives in the prompts. For example, do not say a "faster tempo than the previous part". Instead specify the BPM values. 5. Do NOT make Jazz and Classical music. 6. avoid slow or soft music. 7. keep each part no longer than 30 seconds and no shorter than 20 seconds. 8. Use the form to make the music engaging and interesting, as each part by itself is not that interesting. 9. Avoid world fusion music. 10. Write the form of each piece first. 11. DO NOT use vague language. Be verbose in the prompts. For example if want more layers in one part, just write exactly what instruments should be as layers. 12. Create contrast through BPM, texture, and instruments between parts. 13. avoid comparative phrases like "added layers" and "fuller sound". Instead be verbose, be precise and add absolute phrases only. 14. To repeat a part with variation in the chosen music form, reference that original part and in the prompt for the new part mention what changed. Write the final output in the json code format like this: { PART_NUMBER: ["PROMPT", LENGTH_IN_SECONDS, REFRENCED_PART], PART_NUMBER: ["PROMPT", LENGTH_IN_SECONDS, REFRENCED_PART], ... } Create 10 such pieces. Each of the 10 description is followed by the corresponding json snippet.
\end{cmhlisting}

A sample response from ChatGPT is presented in Listing \ref{rpgpt}.

\begin{cmhlisting}{A sample response from ChatGPT}{rpgpt}
Music Piece 3: "Electro Dance Journey"
Form: A-B-C-B'-A'-C'
Part A: Introduction of the main dance theme
Part B: Development with new elements
Part C: Bridge introducing contrast
Part B': Return of B with variations
Part A': Return of A with changes
Part C': Final contrasting section with a climax.
Prompts and Structure:{  
"1": ["An energetic electro dance track with a driving beat, synth leads, and rhythmic bass. BPM: 130", 25, -1],  
"2": ["A lively electro dance part with additional staccato synth chords and a more complex drum pattern. BPM: 132", 25, -1],  
"3": ["A contrasting electro dance bridge with a softer tone, featuring melodic synth lines and a slower beat. BPM: 125", 20, -1],  
"4": ["Return of the upbeat electro dance theme from Part 2, now with a richer arrangement including layered synths. BPM: 132", 30, 2],  
"5": ["Variation of the initial electro dance theme, integrating elements from previous parts for a fuller sound. BPM: 130", 25, 1],  
"6": ["A climactic electro dance finale, combining the energy of all parts with increased tempo and intensity. BPM: 135", 25, 3]}
\end{cmhlisting}

To evaluate the music generated by our method, we generate 10 samples using our approach and 10 samples using vanilla MusicGen. Each of these samples is 2.5 minutes long. For comparison, we also use 10 samples from Pond5, each exactly 2.5 minutes in length.  and  are two samples generated by our method.  and  are generated by vanilla MusicGen. Using Amazon Mechanical Turk (MTurk), we collect a mean opinion score (MOS) between 1 and 5 for each sample from 10 non-musicians. We ask the human raters to evaluate the overall quality of the music, following the recommended practices in CrowdMOS. For each rating task, the payment amount is calculated to ensure that for an hour of listening, the payment is \$24. The MTurk workers are prequalified by completing 10 rating tasks, with each audio sample being one minute long. Three of the tasks have audio samples with verbal instructions, at a random time, on what score should be chosen. One sample is one minute of silence, and we expect the lowest score of one from the attentive workers. The MOS results are presented in Table 1. The results indicate that adding musical form through our method to MusicGen improves the perceived quality of the music, making it closer to human-composed music.

\begin{table}[]
\caption{presents the mean opinion score (MOS) scores, showing the mean and standard deviation (mean $±$ SD) for each group. Pond5 refers to the music samples taken from the Pond5 website.}
\centering
\begin{tabular}{ll}
  \hline
  \textbf{Model/Source} & \textbf{MOS} \\
  \hline
  Ours & 3.89$±$1.06 \\
  Vanilla MusicGen & 3.50$±$1.08 \\
  Pond5 & 3.98$±$0.81 \\
   \hline
\end{tabular}
\label{tab:myTable}
\end{table}%

The samples generated by MusicGen exhibit a large variation in quality depending on the genre. Some prompts to MusicGen can lead to low-quality samples. In our method, the LLM can sometimes propose prompts that result in low-quality samples. Therefore, the list of sample prompts provided to ChatGPT is narrow in genre and style to avoid proposing MusicGen prompts that lead to low-quality samples. In the evaluation, we use similar prompts or descriptions for both vanilla MusicGen and Pond5 samples to maintain consistency.

\section{Meta Optimization by LLM}

In the previous section, the prompt provided to the LLM for in-context learning is designed manually following a tedious trail and error process. LLMs while understand natural language their interpretation of the input text is sensitive to phrases and words. For instance, chain of thought that asks the LLM to solve problems in multiple steps \citep{wei2022chain} can change the responses of some LLMs. The nuances in the input prompt then motivates exploring the prompt space for a specific task such as generating prompts for a generative music models. There are three downsides to exploring the prompt space manually. First, coming up with a new prompt to evaluate is a daunting and tedious task. Second, in practice a small space can be explored. Third, the LLMs such as GPT are updated regularly and their response changes given the same prompt, requiring the prompt engineering process to be repeated.

\begin{figure}[!ht]
\begin{center}
\includegraphics[width=0.65\textwidth]{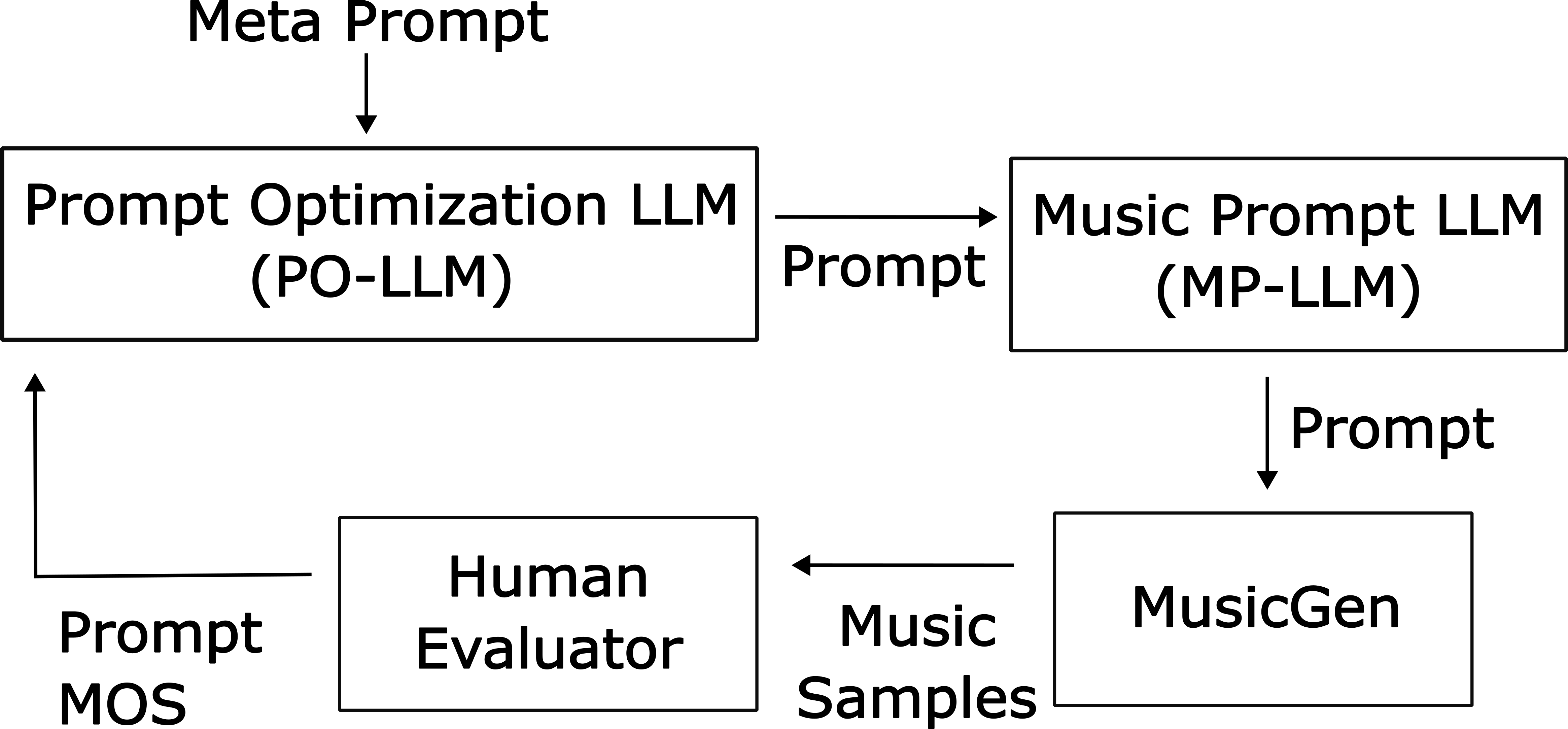}
\caption{The PO-LLM proposes new instructions and few-shot samples for MP-LLM. MP-LLM follows the instructions and generates a set of prompts for MusicGen to generate a coherent music piece with musical form. The generated music is then rated by human evaluators and the average MOS is estimated. PO-LLM is instructed y the meta prompt to consider the previous 5 prompts with highest MOS to propose a new prompt for MP-LLM.}
\label{fig:opti}
\end{center}
\end{figure}

We propose automating the prompt optimization process. The high-level process is illustrated in Figure \ref{fig:opti}. There are two LLMs in this process. The Music Prompt LLM (MP-LLM), as in the previous sections, generates prompts for the generative music model. The Prompt Optimization LLM (PO-LLM) generates new prompts for MP-LLM. The PO-LLM and MP-LLM are both ChatGPT 4 in our experiments in two separate session. However, in general these could be two different LLMs. For the optimization we also need a scoring function. Here we collect MOS for the generated music by the generative music model. Therefore, the PO-LLM attempts to maximize the MOS of the generated music by optimizing the instruction prompt for  the MP-LLM which in turn generates the prompts for the model generating 

 It has been shown LLMs can be used as blackbox optimizers for coefficients in a linear model to text input into another model \citep{yang2023large}. The optimization problem we have here is expected to be harder as instead of optimizing the input to a single model we have two stochastic models chained together and the subjective score is noisy. Our experimental results are promising despite the challenging setup of the optimization problem. The crucial piece is the meta prompt, the prompt that explains the optimization problem to and instructs the PO-LLM.

We use two separate meta prompts, one for each phase of the optimization process. In the exploration phase, from a single seed prompt, which is the best hand engineered prompt, a diverse set of 20 prompts are generated by the PO-LLM. Each of the generated prompts are then used to generate 10 music pieces and each piece is scored by five subjects. The exploration meta prompt is presented in Listing \ref{explore_meta}, and the prompt from the previous section is passed as the sample.

\begin{cmhlisting}{Meta prompt in the exploration phase}{explore_meta}
Assume you are a music composer who is creative and composes in several genres. You want to use a large language model (LLM) to come up with a new composition. The LLM writes a high-level description of each part in the piece. You write a set of prompts for the LLM. A sample prompt is provided to you. You should use this sample to get inspired, do not limit yourself to this, and do your best to be creative to come up with novel and diverse set of instructions. Write 20 separate prompts for the LLM in the JSON format ['_prompt_1','_prompt_2_',..]. Keep the prompts concise. The sample prompt is:
\end{cmhlisting}

In the exploitation phase, the top five prompts with the highest average MOS are passed to the PO-LLM. In each iteration, a new prompt is proposed by the PO-LLM, and it is evaluated. The new prompt is added to the set of top five prompts if the score is higher than that the lowest score in the set. In each iteration, only the top five prompts are retained. After 20 iterations we stop the optimization process. The meta prompt in Listing~\ref{exploit_meta} is passed to the PO-LLM along with the pairs of prompts and corresponding scores:

\begin{cmhlisting}{Meta prompt in the exploitation phase}{exploit_meta}
Assume you are a music composer who is creative and composes in several genres. You want to use a large language model (LLM) to come up with a new composition. The LLM writes a high-level description of each part in the piece. You are given five such prompts with their corresponding scores. You should think of a new prompt that is going to have a higher score than the five samples. The samples projects and their scores are provided in the JSON format: [{'prompt':'_prompt_1','score':_score_1}, {'prompt':'_prompt_2','score':_score_2}, ..} 
\end{cmhlisting}

LLMs are more prone to hallucination when provided a larger prompt \citep{fang2024large}. As the meta prompt can be very large, providing the list of prompt and score pairs in a format such as JSON can help reduce the risk of hallucination. It has been claimed by some studies that the HTML format reduces the chance of hallucination more than the other formats for lists passed to LLMs. This is probably due to the training data of most LLMs containing a large number of HTML documents. For our experiments, we see the JSON format keeps the prompt easy to read and structured well for the LLM without increase the hallucination rate.

\begin{figure}[!ht]
\begin{center}
\includegraphics[width=0.55\textwidth]{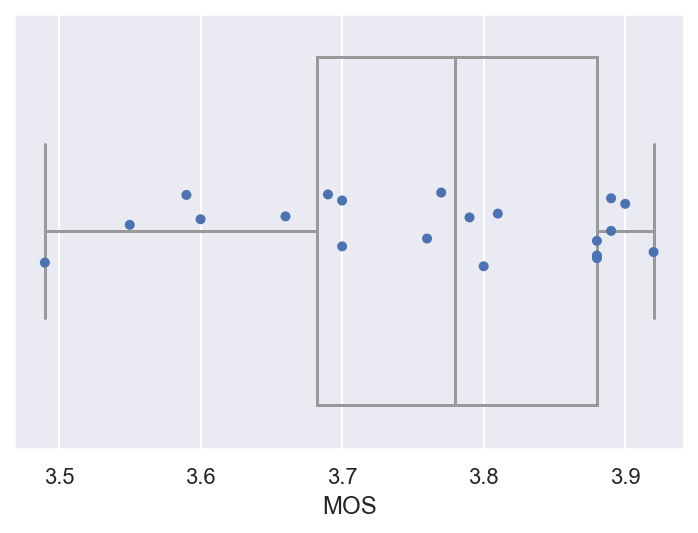}
\caption{The average MOS of the prompts generated in the exploration phase of the optimization method.}
\label{fig:dist}
\end{center}
\end{figure}

Figure \ref{fig:dist} presents the average MOS of the generated music by the prompts generated in the exploration phase. Majority of the scores are lower than the score of the initial prompt. The lower scores are expected, as the initial prompt was tuned manually and probably is close to a local optimum. The spread of scores shows the PO-LLM can generate a diverse set of prompts, which is desirable in the exploration phase.

\begin{figure}[!ht]
\begin{center}
\includegraphics[width=0.65\textwidth]{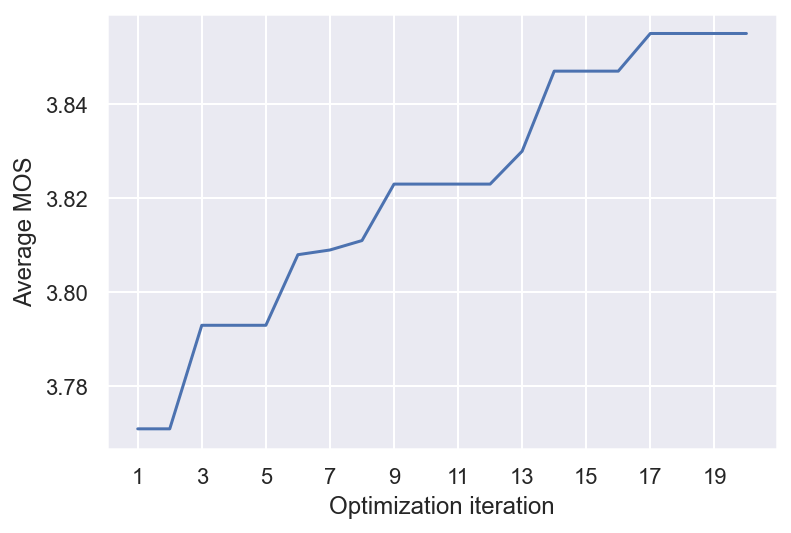}
\caption{The average MOS of the top 5 prompts in each iteration of the optimization.}
\label{fig:prog}
\end{center}
\end{figure}

The average MOS of the top five prompts over optimization iterations in the exploitation phase are presented in Figure \ref{fig:prog}. In more than half of the iterations, the average MOS remains does not change between two iterations due to the MOS of the generated prompt being lower than the top five prompts. The ascend rate and the final to initial average MOS ratio are above the noise level. The prompt with the highest MOS (3.93) at the last iteration is slightly better than the MOS of the initial seed prompt (3.89). The results suggest exploring this method with a non-optimized seed prompt can be a promising research to completely automate the prompt engineering process.

The are other possible approaches to balancing exploration and exploitation in the optimization process. For instance, starting with a larger set of prompts and reducing the list size over iteration can control the transition from exploration to exploitation. The larger the list of prompts and scores in the meta prompt, the more diverse set of prompts the PO-LLM is expected to generate. Our method of two discrete phases is used to reduce the cost of the process as we separately can evaluate the generated prompt in each phase. And collecting MOS for a batch of generated music has a lower time overhead. Therefore, we could more efficiently tune the meta prompt for the exploration phase. The presented results support the possibility of automatic prompt engineering for LLMs generating instructions in natural language for generative music models.

\section{Conclusions}

Generating short music by machine learning models has been the focal of research in the literature, with some notable progress. The generated music by such models lacks structure at the span of a minute or longer. Generating long music with structure has remained an under-researched problem and challenging.  This paper argues due to the nature of the problem, the music generative models cannot learn long scale structure or musical form from a musical corpus.

We then present a novel method to combine music generative models with large language models to generate music withe musical form. The LLM comes up with the form and parts. The LLM communicates with the generative music model in natural language. The technical challenges are discussed and the experiments show the subjects find the generated music as pleasant as the human composed music, based on the MOS. The difference in MOS between our method and the baseline of the generative music model without our method is statistically significant. 

As an extension this paper also presents an optimization method to optimize the prompts for in-context learning maximizing the MOS of the generated music or some other criterion. The limitations of the proposed methods are discussed which are interesting research problems for future work. For instance, continuing motives over multiple parts remains a challenge. In general, generative music models with a greater control over the music aspects is an open problem. Limited size of training data with limited amount of attributes is the main factor. Therefore, future work on larger and richer music data and models with control over more aspects of the music can lead to more useful tool for musicians in practice.

\bibliographystyle{cmj}
\bibliography{cmj}

\end{document}